\documentclass[sigconf]{acmart}%%
\AtBeginDocument{%
  }

\definecolor{supportblue}{RGB}{41,128,185}     % Support and coping mechanisms - calm blue
\definecolor{primaryred}{RGB}{192,57,43}       % Primary stressors - strong red
\definecolor{secondaryorange}{RGB}{230,126,34} % Secondary stressors - orange
\definecolor{emotionalpurple}{RGB}{142,68,173} % Emotional toll - deep purple
\usepackage{makecell}
\usepackage{tabularx,array}
\usepackage{subcaption}
\usepackage{enumitem}
\usepackage{multirow}
\usepackage{longtable}
\copyrightyear{2026}
\acmYear{2026}
\setcopyright{cc}
\setcctype{by-nc-nd}
\acmConference[CHI '26]{Proceedings of the 2026 CHI Conference on Human Factors in Computing Systems}{April 13--17, 2026}{Barcelona, Spain}
\acmBooktitle{Proceedings of the 2026 CHI Conference on Human Factors in Computing Systems (CHI '26), April 13--17, 2026, Barcelona, Spain}
\acmPrice{}
\acmDOI{10.1145/3772318.3791900}
\acmISBN{979-8-4007-2278-3/2026/04}

\begin{document}
%% The "title" command has an optional parameter, allowing the author to define a "short title" to be used in page headers.
\title {Understanding Remote Mental Health Supporters' Help-Seeking in Online Communities
}

\author{Tuan-He Lee}
\email{tl566@cornell.edu}
\orcid{0000-0003-0555-9279}
\author{Gilly Leshed}
\email{gl87@cornell.edu}
\orcid{0000-0002-2308-2825}

\affiliation{
  \institution{Cornell University}
  \city{Ithaca}
  \state{New York}
  \country{USA}
}
\renewcommand{\shortauthors}{Lee et al.}

\begin{abstract} %150 words max
Providing mental health support for loved ones across a geographic distance creates unique challenges for the remote caregivers, who sometimes turn to online communities for peer support. We qualitatively analyzed 522 Reddit threads to understand what drives remote caregivers’ online help-seeking behaviors and the responses they receive from the community. Their purposes of posting included requesting guidance, expressing emotions, and seeking validation. Community responses included providing emotional support, suggesting informational strategies, and sharing personal experiences. While certain themes in posts (emotional toll, monitoring symptoms, and prioritizing caregiver well-being) are shared across remote and non-remote contexts, remote caregivers’ posts surfaced nuanced experiences. For example, they often rely on digital cues, such as voice, to interpret care receivers’ well-being while struggling with digital silence during crises. We discuss the need for supporting communication and information sharing between remote caregivers and receivers, care coordination for crisis management, and design recommendations for caregiver communities.
\end{abstract}




\maketitle

\section{Introduction} % 600 words in total, 400 words for background
Family members and friends play a critical role in supporting individuals with mental health conditions\footnote{In this work, mental illness includes anxiety disorders, mood disorders, and substance use disorders \cite{mcgrath_age_2023}.}, from encouraging adherence to treatment to providing emotional support \cite{sin2012understanding, deane2018carer, lauzier2021caregiver, murnane2018personal}. However, not all supporters\footnote{The terms ``supporter'' and ``caregiver'' are used interchangeably to refer to individuals who engage in supporting their family members and friends with mental illness, as supporters often do not self-identify as caregivers despite providing care to loved ones.} can provide this care in person. Remote supporters, those who provide care across geographic distances, represent an important but understudied group with unique potential to complement care ecosystems. Nearly 6 million Americans serve as remote caregivers \cite{aarp2020caregiving}, yet those who specifically support loved ones with mental illness remain understudied. %While research has examined challenges in remote caregiving generally \cite{bei_barriers_2023, horowitz2018long}, the experiences of remote mental health supporters have received little attention.

Remote mental health caregiving presents distinct challenges that intensify the already significant burden experienced by caregivers. While all mental health supporters face stress, burnout, and frustration \cite{muhlbauer_navigating_2002, estrade_lived_2023, onwumere2017burnout}, remote supporters encounter additional barriers: they cannot directly observe symptoms or behavioral changes, cannot physically intervene during crises, and must rely on limited communication channels that may not convey the full picture of their loved one's mental state \cite{lee2025support}. Geographical distance also limits their access to local support networks and professional resources \cite{bei_barriers_2023, horowitz2018long}. Despite these unique challenges, research has predominantly focused on co-located caregiving relationships, leaving a critical gap in our understanding of how supporters manage mental health caregiving from afar.

Online communities like Reddit have emerged as valuable peer support resources for individuals facing health challenges and caregiving responsibilities \cite{yao2023understanding, progga2023understanding, Homaeian2025community, shoults_analysis_2023}. These platforms offer affordances such as accessibility, anonymity, and asynchronous communication that enable supporters to connect with peers facing similar challenges \cite{tanis2008health}. These features create low-barrier spaces where individuals can candidly share experiences and seek support around sensitive topics \cite{andalibi2021sensemaking, kelly2022exploring}, affordances that may be particularly valuable for mental health caregivers as well. 
Research shows that online forums benefit families facing mental health challenges as they can obtain practical advice and relatable stories from lived experiences \cite{schmutte2023use, Namkoong01022012, siddiqui2023exploring}. However, there is a lack understanding of how remote supporters specifically seek and receive support on these platforms. 
 
To address this gap, we investigate remote mental health supporters' online support-seeking behaviors. Our analysis looks into the kinds of support remote caregivers seek from the online community, the responses they receive from the community, and the caregiving-related topics and challenges they discuss in their posts. Understanding their experiences, concerns, and support-seeking behaviors can inform the design of interventions and digital tools that better support remote mental health caregivers. 
In particular, our study focuses on Reddit, which hosts multiple active mental health support communities and provides publicly accessible discussions that capture natural peer support exchanges. We analyzed Reddit threads (522 original posts and 3,355 comments) from mental health supporter-focused subreddit communities (e.g., r/family\_of\_bipolar, r/depression\_partners).

\newcommand{\RQline}[2]{%
  \par\noindent
  \begingroup
    \setlength{\dimen0}{0pt}%
    \settowidth{\dimen0}{\textbf{#1: } }% width of the label 
    \parshape 2
      2em \dimexpr\linewidth-2em\relax
      \dimexpr2em+\dimen0-0.6em\relax \dimexpr\linewidth-(2em+\dimen0)\relax
    \hspace*{2em}\textbf{#1: }#2\par
  \endgroup

Our paper addresses the following research questions:

\RQline{RQ1}{What are the purposes of remote supporters for posting in mental health caregiving communities?}
\RQline{RQ2}{How do community members respond to remote supporters' posts?}
\RQline{RQ3}{How do remote supporters describe providing care, managing challenges, and coping with stressors when supporting loved ones from a distance?}
}

Following a qualitative thematic analysis of 522 Reddit threads from mental health caregiver communities, we identified four main purposes for seeking online support (RQ1): guidance, emotional expression, validation, and coping strategies. Key topics community members address in their responses to these posts include (RQ2): providing emotional support, offering informational strategies, sharing personal experiences, promoting supporter boundaries, and acknowledging the remote context. These spontaneous discussions about remote caregiving challenges revealed that the topics discussed by remote supporters reflect many common coping mechanisms and stressors found in general mental health caregiving literature (RQ3): they described monitoring symptoms, encouraging treatment adherence, experiencing emotional toll, and prioritizing their own mental well-being. 
However, our analysis also identified challenges and strategies unique to remote caregiving context: remote supporters develop remote monitoring strategies using digital cues, and they struggle with communication disruptions that are particularly critical during mental health crises. Moreover, they often receive community support that is irrelevant as it is not tailored to their distance-specific needs. 

This paper makes the following contributions: (1) identifying the unique strategies and challenges that distinguish remote mental health caregiving experiences from co-located ones, including monitoring and communication, care coordination, and boundary management; {(2) demonstrating how remote supporters use online peer communities for collective sense-making, and the limited guidance they receive given their remote context.

\section{Related Work}

\subsection{Challenges in mental health caregiving} 
% common barriers 
Supporting a family member or a close friend with mental illness creates substantial psychological and physical strain \cite{estrade_lived_2023, onwumere2017burnout, gater2014sometimes}. These supporters must deal with their loved ones' unpredictable symptoms \cite{jeyagurunathan2017psychological} and cope with distress from witnessing their loved one's suffering. The need to remain vigilant to prevent crises and manage symptoms can be emotionally draining \cite{marshall2023caring}. Mental health caregivers often attempt to influence care receivers' decisions and behaviors toward mental health management and recovery \cite{lewis2004conceptualization,lauzier2021caregiver}, from discussing and offering help to expressing negative emotions or guilt \cite{lewis2004conceptualization, butterfield2002health}. 
However, the person being supported may resist or reject help, respond with hostility, or deny the severity of their condition \cite{august2011spouses}, further contributing to the caregiver's emotional strain.

Caring for a loved one with mental illness also often comes with a social price. Many caregivers report reduced social engagement and weakening of their own relationships \cite{azman2019mentally}, and loneliness \cite{guan2023social}. The stigma associated with mental health conditions might further intensify caregivers' burden \cite{van2015stigma}. This stigmatization could exacerbate psychological distress and create barriers to help-seeking, as caregivers fear judgment from mental health professionals \cite{shi2019correlates, mak2008affiliate, vogel2006measuring}.

For remote mental health supporters, some of the challenges may be amplified while new obstacles arise as a result of the geographic distance. Research on long-distance caregiving shows that while remote caregivers serve as vital components of care networks through emotional support, financial assistance, and care coordination \cite{cagle_long-distance_2012,bei_barriers_2023, li2023geographic}, they face substantial barriers that complicate their contributions. 
Remote supporters must rely heavily on local networks to facilitate care coordination and digital communication, and in transnational contexts, they face cultural differences in care expectations and practices \cite{kalavar_im_2020, lee_caring_2015, sethi_caregiving_2022}.
The mental health context may present additional complications for remote communication, particularly in interpreting emotions and establishing emotional connections through digital communication, making it challenging for caregivers to assess their loved ones' mental state and provide appropriate support \cite{lee2025support}. How remote caregivers interact with and influence receivers' mental health management from a distance remains unclear. Research on general long-distance caregiving indicates that remote supporters experience guilt, helplessness, and distance-related distress \cite{lee_caring_2015, horowitz2018long}, while some studies suggest elevated caregiver burden and social isolation among remote caregivers \cite{bei_barriers_2023, li2023geographic}.

\subsection{Online communities for family/friend caregivers}
For people experiencing mental illnesses, online communities serve as an avenue for social and informational support \cite{progga2023understanding, de2014mental, milton2023see}. Less studied are the mental health caregivers who, given reduced social engagement \cite{azman2019mentally} and mental health stigma \cite{van2015stigma}, frequently turn to online forums and communities to seek advice, express emotional distress, and connect with peers \cite{huang2025online, smith2019social}. 
While studies have increasingly explored social media platform use among caregivers and patients in other contexts such as cancer or elder care \cite{hamm2013social,johnson2022s, cooper2021caregiver, shoults_analysis_2023}, research specifically examining online help-seeking behaviors of mental health family caregivers remains particularly scarce. 

% benefits of online social support for caregivers
Online communities provide non-judgmental space where caregivers can openly disclose challenges without fear of social repercussions \cite{daynes2023online}. Such connections create a sense of bonding among caregivers with positive effects including improved coping strategies \cite{Namkoong01022012}, reduced caregiver burden \cite{scheid_social_2009}, decreased isolation \cite{daynes2023online, johnson2022s}, and enhanced psychological resilience \cite {lok2021relationship}. The anonymity of online health forums \cite{tanis2008health} is particularly valuable for those experiencing stigmatized conditions, encouraging caregivers to speak freely, ask sensitive questions. %Receiving informational and emotional support in such spaces can improve caregivers' subjective well-being, and helps release stress \cite{yen2023building}. 
Online peer networks can be especially empowering for mental health caregivers where formal services are scarce \cite{siddiqui2023exploring}.

%\subsubsection{Challenges and considerations in online support-seeking for caregivers}
At the same time, family caregivers face unique challenges when seeking support online, including navigating relational boundaries and privacy concerns when discussing sensitive issues in online spaces \cite{yamashita2018information}. Caregivers must carefully consider what to share publicly, balancing the benefits of sharing illness and soliciting support against protecting their loved ones' privacy and avoiding oversharing sensitive details \cite{johnson2022s}.

% lack of professional oversight
While online peer support offers caregivers mutual help, the literature notes several limitations to its effectiveness.
Research indicates that peer groups without clinical supervision may disseminate inaccurate or potentially harmful advice when addressing critical issues such as medication management, crisis intervention, or legal concerns \cite{wadden2021effect}. Trained or professional moderators are essential for ensuring safe, accurate guidance and fostering trust among users \cite{marshall2024understanding}. 
% inadequate response
Another limitation derives from the inadequate and inconsistent response when mental health caregivers seek help online. Many forums suffer from low peer response rates, often due to volunteer burnout or lack of engagement infrastructure. Studies of online interventions for family caregivers consistently document engagement failures \cite{atefi2024adherence} and the challenge of sustaining engagement \cite{lederman_support_2019}. Researchers found that moderated online mental health conversations yielded higher engagement and improved emotional support compared to unmoderated ones \cite{wadden2021effect}. While peer networks can foster empathy and shared understanding, the lack of professional oversight and inconsistent engagement significantly limit their utility for caregivers facing complex mental health challenges. These findings underscore the need for understanding the dynamics of peer support to ensure sustained participation and effective social support.

\subsection{Research gap and motivation}
Our review shows that family caregivers face substantial challenges that drive them to seek online support, and the online communities they turn to may offer both significant benefits and notable limitations. We lack a clear understanding of the specific kinds of support that remote mental health caregivers seek in these online communities, and what they discuss in these forums. Furthermore, it is unclear what types of support remote caregivers actually receive in response to their posts, and whether these responses align with their expressed needs. To bridge these gaps, we examine remote caregivers' discussions on a public social media platform to understand the ways in which remote supporters engage in seeking online peer support, and how community members respond to them.

\section{Method}
Given our interest in understanding online discussions about remote mental health support, we considered various online forums such as Reddit, Facebook Groups and Discord. %Instead of asking caregivers to report on their online interactions, through an interview or survey study, 
We decided to analyze existing social media discussions as natural representations of people's actual thoughts and communication patterns \cite{gui2017investigating}, and reflect spontaneous and authentic online peer interactions.
We focus the analysis on Reddit because it hosts communities for caregivers of a variety of mental health conditions. Research suggests that the pseudonymous nature and the use of throwaway accounts on Reddit foster the disclosure of stigmatized or personal experiences, enabling more genuine self-expression \cite{de2014mental, andalibi2016understanding}. Reddit-based studies have demonstrated success in examining sensitive topics by accessing discussions that individuals might avoid in more identifiable contexts \cite{gamage2022deepfakes}. In this section, we describe our process of collecting, processing, and analyzing Reddit discussions related to remote mental health support. The data collection and analytic process is visualized in Figure \ref{fig:method}.

\begin{figure*}
    \centering
    \includegraphics[width=1\linewidth]{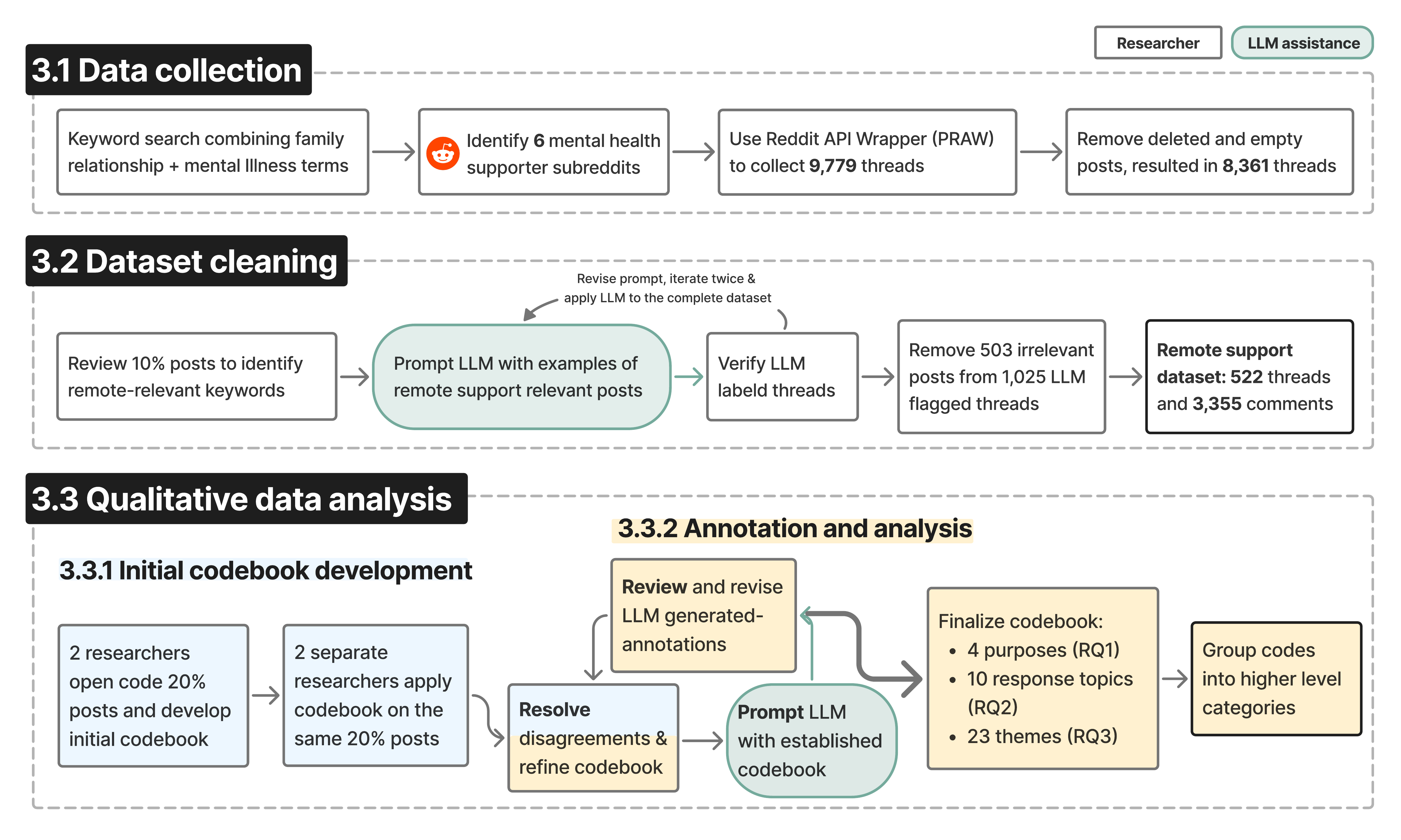}
    \caption{Data collection and qualitative analytic process}
    \label{fig:method}
\end{figure*}

\subsection{Data collection}
%\subsubsection{Subreddit data collection}
Relevant mental health supporter communities were identified via Google Search (Incognito mode) and Reddit, using keywords \cite{carik2025exploring, gauthier2022will} that combined family relationships (e.g., “family member,” “partner”) with mental health conditions (e.g., depression, anxiety, bipolar disorder, addiction, PTSD, eating disorders, and general terms). We also reviewed the recommended subreddit lists and resources provided in the sidebars of identified communities to expand our sample of relevant support forums.
We selected six mental health supporter-specific subreddits based on community size (minimum 5,000 members) and diversity of mental health conditions represented. Larger communities were prioritized to ensure adequate representation of diverse remote caregiving experiences and sufficient data volume for our analysis. The selected subreddits included: r/depression\_partners (13K users), r/family\_of\_bipolar (9.3K users), r/BipolarSOs\footnote{SOs refers to significant others.} (48K users), r/SchizoFamilies(5.3K users), r/AlAnon\footnote{AlAnon refers to those affected by someone else's drinking.} (86K users), and r/AdultChildren\footnote{"Adult Children" of Alcoholics refers to those raised by alcoholic or otherwise dysfunctional caregivers.} (65K users).  %We excluded smaller communities such as r/raisedbybipolar (4.7K users) and r/CPTSDpartners (4.6K users). 

Using Python Reddit API Wrapper (PRAW), we collected data from these communities between August 2024 and January 2025. For each post, we collected post ID, title, date, URL, original content, comments, and upvote count. Due to API limitations that restrict access to the most recent 1,000 posts per subreddit, we continue collecting new posts during this timeframe to maximize data coverage. 
The resulting dataset included posts spanning from October 2019 to January 2025. We initially retrieved 9,779 threads (including original posts and comments) from the six subreddits. After removing posts deleted by the original posters and posts without textual content (attachments only), our dataset comprised of 8,361 threads (see Table \ref{tab:datasets}).

\subsection{Dataset cleaning} 
To identify which posts in the dataset are relevant to remote mental health caregiving, we adopted a human-LLM hybrid technique. We manually reviewed a subset of posts (10\%) to identify keywords relevant to remote support giving. Two researchers independently coded posts for phrases explicitly indicating physical distance (e.g., live in different states) or implicitly suggesting separation (e.g., video calls), then compiled results. The common keywords and phrases indicating geographic distance fell into four categories: location/separation (e.g., long distance relationship/LDR, different state(s), another country, abroad/overseas, live far away); distance measurements (e.g., hours away, miles away); changes in living arrangements (e.g., moved out, moved away, live separately); and remote communication methods (e.g., video call, Facetime). Following previous studies' approach \cite{namvarpour2024uncovering, zhang2025dark, pang2025understanding}, we crafted a prompt that included one exemplary post from the manually labeled posts. We instructed GPT-4 to follow the same procedure (label yes, no) and provide keywords used for identification. 

To validate LLM identification, we conducted two rounds of manual review to refine our prompts, comparing GPT-4's classifications against our manual annotations in each round. In the first round (15\% of the dataset), GPT identified 11.5\% of posts as relevant, with 29.8\% precision and 94.2\% recall. We then refined the instructions to exclude irrelevant cases, such as posts describing emotional distance rather than geographical distance, or where keywords described situations like ``work remotely.'' We then tested the refined prompts on another subset of posts (20\% of the dataset), and the second round achieved 45.9\% precision and 95.8\% recall, with 10.8\% of posts flagged as relevant. Given the high recall rates (>94\%) but variable precision, we prioritized recall to avoid missing relevant posts in this exploratory study. We applied the final refined GPT prompts to the complete dataset and manually reviewed all GPT-identified posts to eliminate false positives.

In total, 1,025 threads (13.6\%) were flagged by LLM as relevant. We read the title and content of every LLM-identified post to ensure it met our inclusion criteria: the post indicates that the poster is a family member, partner or friend who shows care and concern for a loved one with mental illness (i.e. the care receiver) and describes an existing close tie with the receiver, the poster is not co-located with the care receiver, and the post indicates intention to provide or actively engages in remote caregiving. Through this filtering process, we removed 503 threads, resulting in 522 threads focused on remote mental health support for final analysis. The ratio of remote posts versus non-remote posts ranged from 6\%-11\% across the six subreddits (Table \ref{tab:datasets}). 

% confirm again
\begin{table*}[htbp]
  \caption{The dataset included threads collected from six subreddits, then being identified as relevant to remote mental health support.}
  \label{tab:datasets}
  \begin{tabular}{lccc}
    \toprule
    \textbf{Subreddit} & \textbf{Collected threads} & \textbf{Relevant to remote support} & \textbf{\%}\\
    \midrule
    r/depression\_partners & 1102 & 66 & 6\%\\
    r/family\_of\_bipolar & 1310 & 144 & 11\%\\
    r/BipolarSOs & 2328 & 85 & 4\% \\
    r/SchizoFamilies & 775 & 50 & 6\% \\ 
    r/AlAnon & 1633 & 93 & 6\% \\
    r/AdultChildren & 1213 & 84 & 7\%\\
    \midrule
    Total & 8361 & 522 & 6\% \\
    \bottomrule
  \end{tabular}
\end{table*}

\subsection{Qualitative data analysis}

\subsubsection{Initial codebook development}
The first author and one research assistant independently open-coded a 20\% sample \cite{saldana2021coding}, by reading the posts and their corresponding comments and attaching descriptive labels to identify initial meanings. We then reviewed, discussed and aligned our codes to establish a shared understanding toward creating a codebook. Through this line-by-line coding, we recognized that posts included one or more purposes, i.e., what they were seeking to obtain through the post, and that posts and responses covered various topics or themes. We identified 4 purposes, 10 response topics, and 18 themes in original posts. Some examples of the themes include \textit{monitor symptoms and treatment adherence}, \textit{limit contact to protect well-being}, \textit{receiver avoid communication.} Examples of response topics include \textit{provide companion support}, \textit{share personal experience} and \textit{emphasize self-care}. 

Two separate researcher assistants then applied the initial codebook to the same 20\% posts in Atlas.ti. Following a collaborative qualitative data analysis approach \cite{richards2018practical, cornish2014collaborative}, the team met to discuss disagreements in the coding and revised the initial codebook to improve the qualitative reliability of our annotations \cite{mcdonald2019reliability}. 

\subsubsection{Annotation and analysis}
We implemented GPT-4.1 to assist with the annotation and support the iterative process of theme development following recent studies that use LLM to filter relevant content in large-scale datasets \cite{vakeva2025don, lloyd2025ai}. Specifically, we used LLM-assisted coding to efficiently apply new codes across the dataset as our codebook evolved. When new themes emerged during weekly team discussions, LLM could quickly identify potential instances across all 522 posts and 3,355 comments for human review. We maintained annotation quality through human-developed codebooks, line-by-line researcher review of all content, with LLM serving as an annotation assistant rather than replacing human judgment \cite{gao2024collabcoder}. 

\textbf{Integrating LLM in the workflow:} We exported an unannotated QDPX file from Atlas.ti, and used a Python script to instruct GPT-4.1 to annotate posting purposes, post topics, and response topics for each post and comment using the codebook. The output was an LLM-annotated QDPX file that we then imported back into Atlas.ti for verification and further analysis. 
We tracked annotations in a spreadsheet with rows for each of the 522 posts and 3,355 comments and columns for each code. The spreadsheet allowed us to efficiently compare annotations between multiple researchers and the LLM to identify and resolve disagreements through discussion.

\textbf{Review and validation}: All LLM-generated codes were reviewed by the research team. 
During the initial review of 20\% of the posts, we found that the LLM annotations aligned well with codes describing specific actions or events  (e.g., \textit{receiver being institutionalized}) but required more revision for codes requiring interpretation of context or relational dynamics (e.g., \textit{rejection or avoidance of care}).
In response, we refined the LLM annotation prompts through three iterations to reduce discrepancies between human and GPT-4 annotations \cite{choksi2024under}. At each iteration, we applied updated instructions to the entire dataset, and divided the dataset so that two researchers independently reviewed and overrode LLM annotations as needed. Throughout this process (May--July 2025), the team met weekly to discuss and resolve disagreements, revise the codebook, and make final coding decisions through consensus.

\textbf{Finalizing the codebook:} 
We conducted multiple rounds of collaborative coding and codebook review until theoretical saturation was reached and no new codes emerged from the data. Finalizing the codebook included: (1) identifying new codes and added them to the codebook, for example, \textit{supporter seek formal support services} and \textit{illness impact on relationship}; and (2) grouping codes of post themes and response topics into categories. A list of codes and their definitions is included in Appendix ~\ref {codebook}.

In the final codebook, each post was coded with 1-2 \textbf{purposes} and one or more \textbf{themes}, and each comment was coded with one or more \textbf{response topics}:

\begin{itemize}
    \item \textbf{Purpose} was applied to each post, recognizing the main motives for remote supporters to post (RQ1). We followed previous approaches \cite{ibrahim2024understanding} and assigned 1-2 purposes to each post out of 4 purposes (Table \ref{tab:codebook_purpose} in the Appendix). 

    \item \textbf{Response topic} was applied to each comment, recognizing the community reactions to the post (RQ2). We grouped the 10 codes into five categories: provide emotional support, offer informational strategies, share personal experience, promote supporter well-being and boundaries, and acknowledge the remote support context  (Table~\ref{tab:codebook_response} in the Appendix). 

    \item \textbf{Theme} was also applied to each post, recognizing what the remote supporters described in their posts (RQ3). We drew on Stress Processing Theory \cite{pearlin1990caregiving} to organize the stressors we identified into 8 primary stressors (direct demands arising from the care context), 5 secondary stressors (spillover strains affecting other life domains), and 1 outcome (consequences for the caregiver’s well-being) (Table~\ref{tab:codebook_theme_partA}, \ref{tab:codebook_theme_partB}, \ref{tab:codebook_theme_partC} in the Appendix). 
  \end{itemize}

\subsection{Quantitative analysis and measures}
To further address RQ2 (How do community members respond to remote supporters’ posts?), we complemented qualitative analysis with statistical analyses exploring relationships between post characteristics and community responses. These analyses examined qualitative (response topics) and quantitative engagement metrics (e.g., upvotes, number of comments, average comment length). Given the non-normal, skewed distributions typical of social media data \cite{horne2017identifying, yu2024characterizing}, we used non-parametric Mann-Whitney U tests for group comparisons. 
All analyses were conducted using Python 3.8:
\begin{itemize}
    \item \textit{Remote vs. non-remote posts comparison:} To better understand how posts about remote support fit within the broader mental health supporter communities, we compared engagement metrics (upvotes, comments per thread, post length, comment length) between the 522 remote support posts and the rest of the dataset (7,841 posts that didn't mention remote support). We used Mann-Whitney U tests for these comparisons and applied Holm-Bonferroni correction across the four metric comparisons (Table \ref{tab:remote_nonremote}).
    \item \textit{Post metrics by purpose:} We examined the relationship between the post's purpose and engagement metrics: number of post upvotes, number of comments per thread, and average comment length per thread (word count). This analysis assessed whether certain purposes generated different community responses. We used Mann-Whitney U tests to compare posts with vs. without each purpose on the engagement metrics, applying Holm-Bonferroni correction across the four purpose comparisons within each metric (Table \ref{tab:purpose_metric}).
    \item \textit{Response topic by purpose:} We further analyzed whether posts received responses aligned with their expressed support needs. For each posting purpose, we calculated the percentage of response topics that appeared in the comments (Table \ref{tab:purpose_response}). We used chi-square tests to examine if the distribution of response topics differed significantly across purposes.
\end{itemize}

\subsection{Data collection and analysis limitations}
Our approach of collecting and analyzing data has limitations. First, we focused on six subreddits covering five mental health conditions and may have overlooked relevant communities, particularly smaller specialized support groups with different characteristics.
Second, we used LLM assistance to identify relevant threads with $\approx$5\% false negative rate ($\approx$95\% recall). While we prioritized recall over precision due to low prevalence of remote caregiving posts and manually verified all LLM-identified posts, some relevant threads may have been missed if they didn't match the LLM's learned patterns. We also applied LLM annotation for qualitative analysis while maintaining human oversight through collaborative coding and established codebooks. Although this approach risks over-reliance on LLM outputs, the enhanced annotation capacity outweighed this concern. We intentionally avoided fully automating LLM annotation, as studies show LLMs often capture superficial patterns rather than nuances and emotional complexity in mental health contexts \cite{xu2024mental, feuston2021putting}.
Finally, our thematic analysis approach focused on understanding experiences rather than quantifying linguistic patterns, may limit insights into specific communication differences between remote and co-located caregivers.

\subsection{Ethical considerations}
Our University IRB exempted this study from review as it was determined not to involve human subjects research. In spite of that, we took precautionary steps to protect the privacy of community members, following HCI and CSCW research's practices \cite{fiesler2024remember, chancellor2019human}. %We did not collect usernames, as our goal is to identify broader patterns in how supporters navigate remote caregiving rather than focusing on individual stories. 
In our findings, we lightly paraphrase quotes to reduce searchability while preserving their original meaning, with careful review to maintain authenticity. To protect posters' anonymity, we verified that the reworded quotations could not be linked back to the original source by searching for them on Google and Reddit \cite{fiesler2024remember}. 
We believe this research provides insights into the experiences and challenges of remote mental health supporters. These measures balance the benefits of understanding remote caregiving challenges with appropriate protection of user privacy and well-being.

\section{Findings}
Our dataset included 522 posts and 3,355 comments from six mental health support subreddit communities, with posts averaging 383 words and a mean of 8.69 upvotes (Table \ref{tab:summary_stats}). Each post generated an average of 6.43 comments, with comments averaging 86 words in length. %The wide standard deviations and the difference between mean and median values across all metrics suggest considerable variation in engagement.
To contextualize the characteristics of remote support posts within general community posts, we compared post length (in word count) between posts that were specified as remote (522 posts) vs. the rest of the database, where remote support was not specified (7,841 posts). The \textit{Length of original post} of remote support posts was significantly longer (median = 318 words) than other posts (median = 208 words) ($p<.001$, $g=0.408$, medium effect). It is possible that because posts that discuss remote support represent a minority in the community (6\% of posts), posters need to provide additional context for their posts, making them longer. Detailed engagement metrics are discussed in Section  \ref{response_metrics}.

We now present the findings from our dataset analysis, on purposes of posting (RQ1), the responses from community members (RQ2), and the topics discussed in original posts (RQ3). All the quotes were extracted from Reddit threads, edited, or paraphrased to ensure anonymity. We attribute quotations using the source subreddit's target relationship and condition (e.g., partner, depression) to indicate community context; however, individual posters may not necessarily align with the subreddit's intended audience.

\begin{table*}[h]
\caption{Summary statistics for 522 posts and 3,355 comments. Number of comments and comment length are calculated per thread.}
\label{tab:summary_stats}
\small
\begin{tabularx}{\textwidth}{lXXXX}
\toprule
 & \textbf{Length of post (words)} & \textbf{Upvotes of original post} & \textbf{Number of comments} & \textbf{Length of comment (words)} \\
\midrule
\textbf{Mean (Std)}        & 383.32 (258.25) & 8.69 (13.66)  & 6.43 (6.76) & 86.29 (104.42) \\
\textbf{Median (Q1-Q3)}    & 318 (214–486) & 5 (3–10) &  4 (2–8) &  57 (24–108) \\
\bottomrule
\end{tabularx}
\end{table*}

\subsection{Purposes of posting}
In response to the first research question, \textbf{what are the purposes of remote supporters for posting in mental health caregiving communities?}, we identified four primary purposes across 522 posts: request guidance on supporting and understanding care receivers, express emotion and share caregiving experience, seek validation and shared experience, and seek coping strategies for oneself. Table \ref{tab:codebook_purpose} in Appendix provides an overview of purposes, their frequencies, and paraphrased example quotes.

\subsubsection{Request guidance on supporting and understanding care receivers (55\%)} \label{guidance}
The majority of posts were seeking advice from the community on supporting a loved one with mental health challenges or understanding their behaviors. Specifically, they requested guidance on communication strategies, methods for comforting or motivating the receiver, and approaches for encouraging receivers seek professional help. 
Supporters also sought help understanding confusing behaviors or trying to interpret whether behaviors were symptom-related: \textit{``Now that she wants to break up again, I can't tell if it's an episode because we're long-distance. I’m so worried and confused if it is an episode. Is there ever a way to know?''} (family, bipolar).
Some supporters struggled with the paradox that their well-intentioned communication efforts, meant to show their care and concern, could inadvertently burden the care receiver. This left supporters feeling lost about how to proceed, as one poster described: \textit{``He admitted my texts sometimes stress him out, but when I offered to stop, he reassured me to continue. How do I be a present, supportive partner?''} (partner, depression).

\subsubsection{Express emotion and sharing caregiving experiences (34\%)}  \label{emotion} 
Many posters use these platforms as outlets to express emotional exhaustion and share their personal caregiving journeys. Posters typically describe their feelings, including worry, exhaustion, confusion, helplessness, loneliness, and guilt, aligning with previous literature \cite{siddiqui2023exploring, estrade_lived_2023, azman2019mentally, muhlbauer_navigating_2002}. Adding to prior work, we identified posts in this category that described the mindset and journey of prioritizing one's own well-being and establishing boundaries due to caregiver burnout. Posts also reflected on the impact of the illness on their relationships with care receivers. For instance, one poster expressed: \textit{``I'm devastated about losing this friend - had to step away''} (family, bipolar). Another poster described a similar sentiment toward their sister: \textit{``I had to protect myself from all that toxic energy, but it breaks my heart that I've lost the sister I care so much about.''} (family, alcohol use).

\subsubsection{Seek validation and shared experience (25\%)}  \label{validation} 
Posters also frequently seek shared experiences to understand and validate their own situations and feelings. Posters are not necessarily seeking advice, but rather confirmation and reassurance from peers who can relate to their experiences. A typical expression for posts in this category is: \textit{``Has anyone been through a similar situation?''}. 
Some posters describe turning to peer knowledge due to frustration with finding relevant and useful information elsewhere. One poster expressed: \textit{``It's getting really tough to keep having my love and care pushed away, and I need to hear from people who've been through this or might understand what's going on.''} (partner, depression). Some also specifically sought success stories and hope for positive outcomes, as reflected in questions like \textit{``Has anyone actually succeeded in helping their loved one achieve long-term independence and stability?''} (family, schizophrenia).

\subsubsection{Seek coping strategies for oneself  (17\%)}  \label{coping} 
Some posters explicitly sought strategies to manage their own well-being while in a caregiving role, such as setting boundaries, accepting their limitations, and deciding whether to continue providing support. This category focuses specifically on the caregiver's own emotional and psychological needs, where the poster explicitly requested advice about coping for themselves instead of caregiving for their loved one. These requests covered various aspects of self-care, from establishing boundaries such as \textit{``What's the best way to set boundary with my alcoholic sibling who's planning to visit me overseas?''} (family, alcohol use) to accepting their limitations as caregivers, asking questions like \textit{``Can I really help my alcoholic sister who's spiraling, is there actually anything I can do to help her, and if not, how do I come to terms with that?''} (family, alcohol use). Supporters also sought guidance on whether to continue caregiving, questioning the sustainability of their support role: \textit{``Do I wait and give him more time to get better mentally, or should I end things now? I don't want to waste my time or his''} (partner, depression).

\subsection{Community response patterns} \label{reponse_findings}
To answer \textbf{RQ2: How do community members respond to remote supporters' posts?}, we examined response patterns to understand what types of support remote caregivers receive and whether these align with their expressed needs. We identified five categories of response topics: provide emotional support, offer informational strategies, share personal experience, promote supporter well-being and boundaries, and acknowledge the remote support context (for detailed description and paraphrased example quotes, see Table~\ref{tab:codebook_response} in Appendix).

\subsubsection{Provide emotional support (75\%)}

Community responses frequently offered direct empathy, validation, and encouragement. For instance, when posters expressed feeling hurt or rejected by their loved one's behavior, responders often provided reassurance: \textit{``Do not take his sudden distance and coldness personally. He is a different person right now because of his episode. It is not because of anything that you did or didn't do''} (family, bipolar). Members also frequently offered messages of hope and encouragement.
 
Responses also offered \textbf{companionship}, emphasizing that posters were not alone in their struggles. These messages often conveyed shared identity, presence and availability, fostering a sense of belonging. As one member reassured: \textit{``You're normal here! We understand you in a way that few others can.''} Others expressed deep connection through shared experience: \textit{``I have gone through something very similar with a person who has had severe bipolar manic episodes in the past... I could have written this very thing and it has me sobbing''}. Beyond this, community members also extended invitations for ongoing connection beyond the initial post: \textit{``PM me if you'd like to chat''}.

\subsubsection{Offer informational strategies (69\%)}
Peers often provided practical guidance, including \textbf{actionable support-giving advice}, \textbf{educational knowledge}, \textbf{recommendations for external resources}, and \textbf{suggestions of professional help for receivers}. Community members offered specific communication strategies for managing difficult interactions: \textit{``Try not to take anything he says personally and to resist the urge to argue with him. Trying to argue with a delusion just reinforces it. Empathizing with the emotions the person feels about it tends to discharge the tension''} (family, schizophrenia). Beyond communication advice, a common theme was emphasizing that receivers need timely professional treatment before symptoms worsen. %For example, this poster emphasized urgency and provided concrete steps for crisis situations:  \textit{``You need to get him to hospital. Get a mental health team to do an assessment. Or get the police to do a wellness check''} (family, bipolar).
    
When direct intervention wasn't feasible due to distance, commenters guided posters toward local resources: \textit{``I know you feel like your hands are tied because he lives away from you, but there are still ways to support him indirectly. Have you looked into local rehabilitation centers for schizophrenia?''} (family, schizophrenia). In addition, members also shared educational knowledge about mental health conditions, treatments, and medications to help caregivers better understand their situations.  %\textit{``He does not recognize you right now. Bipolar Disorder attacks the prefrontal cortex, and he is probably incapable of feeling anything about your most significant memories together.}'' (family, bipolar).
        
\subsubsection{Promote supporter well-being and boundaries (65\%)}
Community members frequently \textbf{emphasized self-care} and boundary setting as essential for supporters' own well-being. They addressed the guilt that often accompanies boundary setting: \textit{``None of us is capable of managing someone else’s adult life. It’s not your job and it’s not your responsibility. And it won’t be your fault if something happens''}. %(partner, depression).
Members encouraged realistic self-assessment about the sustainability of caregiving efforts: \textit{``Consider the level of work and effort this relationship will take, then honestly assess whether you're prepared for a lifetime of this level of effort''} (partner, depression).  
At the more intensive end, some members \textbf{advised discontinuing support relationships} entirely. This advice was more common in romantic relationships, where members explicitly advised ending the partnership, particularly when the loved one remained unmedicated or unwilling to seek treatment. Advice to discontinue was also more frequent in cases involving substance use disorders. For example, a commenter encouraged the poster to prioritize their own well-being: \textit{``Alcoholism is too much for us. It’s a fight not worth fighting. It’s not our business whether the alcoholic wants to get better or not. Our business is setting boundaries that protect us from the disease. That may mean cutting off financial support or cutting off completely''} (family, alcohol use). 
\subsubsection{Share personal experience (64\%)}
Community members often recounted similar situations to provide both emotional validation and informational insights drawn from their lived experience. Supporters often shared similar struggles and insights: \textit{``That's the hardest part of being a partner to a pwbp (person with bipolar) is that the switch can turn off anytime and they transform into a totally different person. It is important to recognise and understand that this is the bp working''} (partner, bipolar). This experience sharing created reciprocal benefits. When commenters shared their experiences to support others, they often found comfort for themselves through the connection. As one poster reflected: \textit{``It actually makes me feel better hearing that coming from somebody else who has gone through the same type of anguish''} (family, schizophrenia). Another poster expressed: \textit{``I'm so glad to have found this group. Reading some of these stories is like a page from my life over last 7 months''} (family, bipolar). 
    
Notably, individuals with mental health conditions also contributed perspectives from their lived experience as care receivers. One commenter identified themselves as having bipolar disorder and offered advice based on their personal understanding of the condition: \textit{``Bipolar here! Visit him! He may need that connection, but expect he might feel ashamed. When I dissociate, I appreciate just having someone there''} (family, bipolar). 

\subsubsection{Acknowledge remote context (11\%)}
Community members occasionally validated the context of caregiving at a distance. Some responses acknowledged the inherent limitations: \textit{``How are you supposed to monitor their mental health, medication compliance, substance use, and daily routines from a distance?'' }(partner, bipolar) and \textit{``This isn't something you can manage by yourself, especially when you're far apart''} (partner, bipolar). Other responses suggested practical solutions while validating the difficulty. For example, one commenter suggested seeking help from a local support network: \textit{``It must be so hard, having to witness all this at a distance and not being able to physically intervene. Do you know what his local support network is like? Has he got any friends living nearer, any family members who could be supportive? Can you get in touch with any of them?''} (partner, depression).

\subsubsection{Quantitative analysis of community response metrics} \label{response_metrics}

\begin{table*}[h]
\caption{Comparison of Reddit posts that specified remote support vs. the rest of the dataset (did not specific remote support). Metrics are reported as median (Q1-Q3)/ mean (SD). Group differences were assessed using the Mann–Whitney U test with Holm–Bonferroni correction. $p$-values and effect sizes (Hedges' $g$) are shown, with * $p<0.05$, ** $p<0.01$, *** $p<0.001$. }
\label{tab:remote_nonremote}
\centering
\footnotesize
\begin{tabular*}{\textwidth}{@{\extracolsep{\fill}}lllll}
\toprule
\textbf{Metric} & \textbf{Specified as remote support} & \textbf{Not specified as remote support} & \textbf{$p$} & \textbf{$g$} \\
\midrule
$n$ (threads) & 522 & 7{,}841 & -- & -- \\
Length of original post (words) & 318 (214--486) / 383.32 (258.25) & 208 (110--356) / 274.73 (266.31) & $<.001^{***}$ & 0.408 \\
Upvote of original post & 5 (3--10) / 8.69 (13.66) & 7 (3--14) / 12.71 (20.11) & $<.001^{***}$ & 0.203 \\
%Sentiment score & -0.672 (-0.96-0.883)/ -0.16 ± 0.86 & 0 (-0.92-0.879)/ -0.023 ± 0.82 & $<.001^{***}$ & -0.167 \\
Num. comments per thread & 4 (2--8) / 6.43 (6.76) & 5 (2--11) / 8.84 (11.02) & $<.001^{***}$ & 0.223 \\
Length of comments (words) per thread & 76 (54--112) / 94.74 (73.18) & 73 (50--102) / 86.76 (70.10) & $.004^{**}$ & 0.114 \\
\bottomrule
\end{tabular*}
\end{table*}
To provide an additional lens on community engagement (RQ2), we conducted quantitative analysis examining engagement with different post types. First, we compared engagement metrics (upvotes, number of comments, and comment length per thread) between posts that specify remote support (522 threads) and the rest of the dataset (7,841 threads) (Table \ref{tab:remote_nonremote}). We found that remote support posts receive fewer comments (median = 4 vs. 5, $p<.001$, $g=0.223$, small effect). However, comments responding to remote support posts were longer (76 vs. 73 words, $p=.004$, $g=0.114$, small effect). The lower number of responses may suggest that remote supporters receive less immediate community validation and fewer diverse perspectives when seeking help. However, the longer responses to remote posts may indicate that when community members do respond, remote supporters may receive more detailed guidance.

% Reviewer request to tone down 
We also examined community engagement metrics in remote support posts given the post's purpose, comparing posts with and without each purpose (Table \ref{tab:purpose_metric}). Posts \textbf{requesting guidance} received significantly fewer upvotes than those not requesting guidance (median = 4 vs. 7), suggesting these posts receive less community recognition. In contrast, \textbf{expressing emotion} received more upvotes than those not expressing emotion (median = 8 vs. 5) potentially indicating that remote supporters who share their emotional struggles receive stronger peer validation. 

There were no statistically significant differences in upvotes in \textbf{seeking validation} and \textbf{coping strategies}, and no significant differences in the number of comments per thread and average comment length per thread across all purposes.

\newcolumntype{P}{>{\hsize=.75\hsize\raggedright\arraybackslash}X} % Purpose
\newcolumntype{G}{>{\hsize=.45\hsize\raggedright\arraybackslash}X} % Group
\newcolumntype{U}{>{\hsize=1.25\hsize\raggedright\arraybackslash}X} % Upvotes
\newcolumntype{C}{>{\hsize=1.25\hsize\raggedright\arraybackslash}X} % Comments
\newcolumntype{M}{>{\hsize=1.30\hsize\raggedright\arraybackslash}X} % Comment words

\begin{table*}[t]
\caption{\textbf{Post metrics by purpose} (with vs.\ without the purpose) indicated with Median (Q1--Q3) / Mean (SD). $p$-values are reported using Mann--Whitney U test with Holm--Bonferroni correction ($^{*}p{<}.05$, $^{**}p{<}.01$, $^{***}p{<}.001$).}
\label{tab:purpose_metric}

\begingroup
\footnotesize
\setlength{\tabcolsep}{5pt}
\renewcommand{\arraystretch}{1.05}

\begin{tabularx}{\textwidth}{@{}P G | U C M@{}}
\toprule
\textbf{Purpose} &
\textbf{Group} &
\textbf{Post upvotes (count)} &
\textbf{Comments per thread (count)} &
\textbf{Comment length per thread (words)} \\
\midrule

\multirow{2}{=}{Request guidance} &
With    & \textbf{4 (2--7) / 6.05 (6.62)}$^{***}$   & 4 (2--8) / 6.42 (7.35)  & 66 (46--97) / 83.94 (79.98) \\
& Without & 7 (4--14) / 11.96 (18.58)                & 5 (3--8) / 6.45 (5.96)  & 65 (39--89) / 74.79 (56.66) \\
\addlinespace[2pt]

\multirow{2}{=}{Express emotion} &
With    & \textbf{8 (4--14) / 12.43 (18.58)}$^{***}$ & 4 (2--8) / 6.05 (6.66)  & 65 (40--89) / 73.03 (48.56) \\
& Without & 5 (3--7) / 6.79 (9.80)                    & 4 (2--9) / 6.63 (6.81)  & 66 (44--97) / 83.32 (79.32) \\
\addlinespace[2pt]

\multirow{2}{=}{Seek validation} &
With    & 5 (3--10) / 9.40 (14.64)                   & 5 (3--10) / 7.08 (6.28) & 68 (48--105) / 89.88 (87.72) \\
& Without & 5 (3--10) / 8.45 (13.32)                  & 4 (2--8) / 6.21 (6.90)  & 64 (41--89) / 76.48 (63.65) \\
\addlinespace[2pt]

\multirow{2}{=}{Seek coping strategies} &
With    & 5 (3--8) / 8.63 (14.64)                    & 5 (2--10) / 6.85 (6.04) & 68 (45--90) / 78.35 (55.50) \\
& Without & 5 (3--10) / 8.70 (13.46)                  & 4 (2--8) / 6.35 (6.90)  & 65 (42--93) / 80.25 (73.64) \\

\bottomrule
\end{tabularx}
\endgroup
\end{table*}

\begin{table*}[htbp]
\caption{\textbf{Response topic by purpose:} Percentage of posts receiving at least one comment with each response topic, broken down by posting purpose. (\(N\))  = number of posts for each purpose. A post may belong to 1–2 purposes, and a comment may have multiple response topics.}
\label{tab:purpose_response}
\centering
\footnotesize
\setlength{\tabcolsep}{2pt}
\begin{tabularx}{\linewidth}{lc*{5}{>{\centering\arraybackslash}X}}
\toprule
\begin{tabular}[t]{@{}l@{}}Purpose \end{tabular} &
\begin{tabular}[t]{@{}l@{}}Posts\\ with purpose (\(N\)) \end{tabular} &
\multicolumn{5}{c}{Topics in response (\%)} \\
\cmidrule(l){3-7}
 & & Provide emotional support & Offer informational strategies & Share personal experience & Promote supporter boundaries & Acknowledge remote context \\
\midrule
Request guidance & 228 & 86\% & 86\% & 77\% & 76\% & 16\% \\
Express emotion & 162 & 88\% & 73\% & 77\% & 77\% & 14\% \\
Seek validation & 125 & 92\% & 82\% & 84\% & 77\% & 18\% \\
Seek coping strategies & 88  & 90\% & 82\% & 84\% & 82\% & 13\% \\
\bottomrule
\end{tabularx}
\end{table*}

Finally, we examined the specific topics that comments included in response to each post's purpose (Table~\ref{tab:purpose_response}) to determine if the comments reacted to posts in a way that fulfilled their purpose. 
Our analysis indicates some variation in response patterns, though a chi-square test showed no significant association overall  ($\chi^2$(12, N = 2,059) = 3.3, p = .99; Cramér's V = .023). \textbf{Provide emotional support} was the most common topic in responses across all purposes (86--92\%), while \textbf{Acknowledge remote support} was uniformly the least common topic (13--18\%).  This suggests that remote supporters receive relatively consistent types of responses regardless of what they explicitly request. While this means they can reliably expect empathetic support from the community, it may also indicate that their specific needs are not always directly addressed. Remote supporters seeking concrete guidance or validation may find that responses focus more on emotional comfort than practical solutions. The low acknowledgment of remote context is notable, suggesting that distance-specific challenges may be under-addressed even when crucial to the poster's situation.

\subsection{Themes described in caregiving original posts} \label{posttopic}
In this section, we respond to \textbf{RQ3: How do remote supporters describe providing care, managing challenges, and coping with stressors when supporting loved ones from a distance?} 
Remote supporters faced universal caregiving concerns as well as problems that, while not all unique to distance, were significantly complicated by geographic separation. Aligned with previous research on mental health caregiving challenges \cite{lauzier2021caregiver, estrade_lived_2023, onwumere2017burnout, gater2014sometimes}, remote supporters described monitoring receivers' symptoms and treatment adherence, urging receivers to seek professional help, and strategies for prioritizing their own mental health when caregiving became overwhelming. Making these self-preserving decisions was often challenging, as supporters had to overcome guilt and social expectations about their caregiving responsibilities. 
We further identified stressors associated with caregiving, including \textit{primary stressors}---direct challenges from receivers' conditions or behaviors, such as unpredictable receivers' behaviors driven by their symptoms, and \textit{secondary stressors}---relationship and interpersonal impacts, such as being subjected to negative communication from the receiver. We also identified the emotional toll (72\%) that supporters experienced as an outcome of these stressors. Table \ref{tab:codebook_theme_partA}-\ref{tab:codebook_theme_partC} in the Appendix present the themes that were identified in the posts.  

We focus this section on three themes that we found to be unique and shaped by remote caregiving: remote monitoring and communication, care coordination across distance, and setting up boundaries. For each theme, we describe the challenges that remote supports express in their posts, and coping strategies that remote supports discuss as helping address these issues (Table \ref{tab:themes_remote}). 

\begin{table*}[ht]
  \centering
  \caption{Challenges and coping strategies unique to remote context.}
  \begin{tabular}{l|p{1cm}}
    \toprule
    \textbf{Theme / Subtheme} & \textbf{\%} \\
    \midrule
    \ref{digitalmonitoring} \textbf{Remote monitoring and communication} &  \\
    \quad \textit{{Challenges}} & \\
    \quad \quad Receiver avoid communication & 31\% \\
    \quad \quad Rejection or avoidance of caregiver support & 29\% \\
    \quad \quad Information gaps & 7\% \\
    \quad \quad Separation initiated by receiver & 7\% \\
    \quad \textit{Coping strategies} & \\
    \quad \quad Apply multiple digital channels and cues & 21\% \\
    \midrule
    \ref{coordination} \textbf{Care coordination across distance} & \\
    \quad \textit{Challenges} & \\
    \quad \quad Concerns about impact on other family members & 14\% \\
    \quad \quad Conflicts with other supporters & 7\% \\
    \quad \textit{Coping strategies} & \\
    \quad \quad Coordinate with local/other supporters & 13\% \\
    \quad \quad Seek formal support services & 10\% \\    
    \midrule
    \ref{boundary} \textbf{Setting and maintaining boundaries} &  \\
    \quad \textit{Coping strategies} & \\
    \quad \quad Limit contact & 11\% \\
    \quad \quad Leverage physical distance & 5\% \\
    \bottomrule

  \end{tabular}
  \label{tab:themes_remote}
\end{table*}

\subsubsection{Remote monitoring and communication}
\label{digitalmonitoring}
Posters frequently described the \textbf{challenges} of experiencing receivers' avoidance behaviors that disrupted their remote caregiving efforts. This included \textbf{communication avoidance} (31\%) which was particularly problematic during critical moments when communication was most needed. Distance created constraints in communication timing and channels that made it easier for receivers to withdraw from contact when experiencing symptoms or distress. 
Many supporters expressed frustration with one-sided caregiving attempts that felt especially difficult to navigate remotely: \textit{``My long-distance boyfriend is going through depression and requested some time alone. I keep sending messages to let him know I'm still around and thinking of him, but he hasn't replied in over a month''} (partner, depression). 
Even when communication occurred, supporters often felt it was insufficient: \textit{``When I text her, it takes several days to hear from her, and it’s very short. It just feels weird. Like I’m talking to a ghost or half a person''} (family, alcohol use).

%Care rejection
Beyond communication avoidance, supporters reported experiencing \textbf{care rejection} (29\%), when receivers declined or resisted conversations about seeking therapy, taking medication, or discussing their mental health symptoms. This was particularly complex for remote supporters, who lack the ability to assess the situation in person or provide immediate physical presence during difficult moments. For example, one supporter described their frustration with repeated rejections from their parents: \textit{``They refuse to stop drinking and have declined all suggestions for help, including therapy, community support,  moving closer to us so we can at least keep an eye on them''} (family, alcohol use). 

These communication limitations played a role in \textbf{information gaps} (7\%), making it difficult for supporters to assess their loved ones' well-being and safety: \textit{``My question is how can we tell, from afar and with no communication, when he is a danger to himself or others… I'm trying to gauge his state''} (family, bipolar). %Communication breakdowns left supporters without critical information about their loved ones' whereabouts and condition: \textit{`Because of the lack of communication, no one actually knows where he is at the moment''} (family, bipolar). 
Information gaps were especially challenging when \textbf{separation was receiver-initiated} (7\%). When care receivers left home, moved away, or fled due to symptoms, supporters struggled with not knowing their whereabouts. This lack of basic information about location made it difficult to assess safety.
Digital silence was particularly distressing because supporters could not distinguish between normal unavailability and concerning changes: \textit{``His silent periods make me anxious, but so does sending multiple texts quickly and become active on social media. Now I'm anxious because my message hasn't been delivered after almost a full day''} (family, schizophrenia).

Despite these challenges, supporters applied \textbf{coping strategies} by seeking \textbf{multiple digital channels and cues} (21\%) to connect, monitor, and care for their loved ones. They turned to video and voice-based calls for direct observation of symptoms and behaviors, for example, to make sure they takes their medication: \textit{``Every night, she takes her medication over video call with me or one of my siblings, showing us each pill removed from its bottle and followed by food and water''} (family, bipolar). Supporters also learned to detect mood changes, energy levels, and behavior changes through audio cues and patterns during calls. For example, this supporter detected substance use through speech patterns: \textit{``We live long distance and he doesn't drink around me. I can tell he's been drinking at least weekly because his speech is slurred when we talk over FaceTime''} (family, alcohol use). 

Beyond direct calls, supporters monitored digital signals such as text messages, social media activity, and location data for signs of mental state changes.  For instance, some posters view heightened social media activity as a signal of maniac status: \textit{``The only way we were seeing if he is ok, was through his manic Instagram stories... We were telling ourselves at least he is safe''} (family, schizophrenia).  During crises, technology may heighten their digital monitoring, including location tracking: \textit{``I end up calling him over 10 times, telling him not to hurt himself, and checking his location every minute''} (partner, depression).

\subsubsection{Care coordination across distance} \label{coordination}

Coordinating among multiple caregivers and healthcare providers often involves \textbf{challenges} for remote supporters. They raised \textbf{concerns about the impact on other family members} (14\%), who are living with the receiver, and were worried about how to better support them. For instance, one supporter sought help for their mother who was also caregiving: \textit{``What can I do to support my mom in supporting my sister so she feels more equipped and not just get completely consumed by this nightmare''} (family, bipolar). 
Remote supporters also provided emotional support to other family members who were dealing with the situation directly: \textit{``I called my father to comfort him because my mom's mental state really overwhelms him''} (family, alcohol use).
In other situations, posters mentioned \textbf{conflicts with other supporters} (7\%), particularly disagreeing about treatment or care priorities and decisions: \textit{``He's staying with his parents now and has mostly come back to reality, but they convinced him to stop taking his prescribed medication''} (family, bipolar).  
When \textbf{receivers were institutionalized }(14\%) in psychiatric facilities or hospitals, some remote supporters found that hospital or facility staff did not adequately work with family members to provide updates or incorporate their knowledge of the care receiver's history. 

Remote supporters described \textbf{coordinating with local supporters} (13\%) as a \textbf{coping strategy} for remote care coordination, aligned with previous research on long-distance caregiving. \cite{cagle_long-distance_2012, bei_barriers_2023, li2023geographic}. They collaborated with family members and friends who lived closer to the care receiver and could provide in-person care and support. %For example, one poster concerned about her sister relied on their parents for updates and monitoring: \textit{``She lives with my mom about an hour away, so I'm not around her much, but my mom has noticed things like her not sleeping well and obsessing over past trauma''} \textcolor{red}{(add subreddit)}.  
For example, one poster relied on her sister's partner for updates: \textit{``Her boyfriend calls me every day with updates, and he says the hospital psychologist is working to get her on a consistent medication routine''} (family, schizophrenia).

Remote supporters also reported \textbf{seeking formal support services} (10\%), specifically when the situation became too difficult to handle on their own or required outside intervention. 
These services included mental health professionals, emergency services (e.g., mobile crisis units, paramedics), law enforcement, and crisis hotlines (governmental and non-governmental organizations). They also sought social services such as shelter services, or home assistance when their loved one needed basic resources or posed safety concerns to family members. When supporters lacked information or access to provide direct support, some contacted emergency services or requested wellness checks. For instance, during emergency situations, participants requested police wellness checks: \textit{``I have talked him out 5 or more suicide attempts where recently I had to call the cops on him.''} (partner, depression). 
Some participants described how they weren't always sure about when or how to get outside help:  \textit{``I've been thinking about calling 988 \footnote{988 is the number of The National Suicide Prevention Lifeline in the U.S.} but I don't know if that's the right thing to do or if it will make things worse''} (family, schizophrenia). 

\subsubsection{Setting and maintaining boundaries} \label{boundary} 
Unlike in-person mental health support, where the all-consuming nature of caregiving leaves little room for the caregiver to attend to their own well-being \cite{lavers2022systematic, karp_mental_2000, broady_how_2015}, for remote supporters, the distance offered \textbf{coping strategies} for setting and maintaining boundaries. 
To protect themselves from emotional turmoil, hostile interactions with the receiver, or feelings of rejection, the remote supporters set boundaries for their caregiving efforts to protect their well-being.

Some posters took concrete steps to \textbf{limit contact} (11\%) with the receiver. These decisions were sometimes described as coming after years of negative interactions: \textit{``After putting up with decades of being constantly criticized—I'm talking 30 years of this—I finally reached my breaking point last year. Now we barely talk''} (family, bipolar). Others made more definitive choices when emotional harm became intolerable: \textit{``Communication between us has stopped, 2 weeks ago because I got tired of the verbal abuse''} (family, bipolar). 

A few posters described \textbf{leveraging the physical distance} (5\%) as an intentional coping strategy to preserve their well-being. Some maintained pre-existing geographic separation, while some actively left shared living situations. For example, one poster explains how they intentionally create distance: \textit{``I intentionally live far away to prioritize my own family. I do not want to be an arm’s reach away to fix a problem (or get blamed for not showing up)''} (family, alcohol use). Similarly, another poster viewed the physical distance as beneficial: `\textit{`Through my own years of counseling and physical distance, I feel that I am able to be the navigator for this'' (family, schizophrenia)}. These quotes illustrate how some supporters used distance for self-protection rather than viewing it as a caregiving obstacle.

\section{Discussion} 

In this study, we examined remote mental health supporters' online support-seeking behaviors and community interactions through analysis of posts and responses in Reddit mental health communities. In their posts, they discuss challenges---monitoring well-being from a distance, coordinating crisis-oriented (rather than preventative) care, and managing boundaries while coping with guilt---consistent with recent research on remote mental health supporters and distance caregivers more broadly \cite{lee2025support, horowitz2018long}. Our findings further show that to address these challenges, remote supporters engage in collective sense-making with online peers---seeking help to interpret receivers' behaviors, respond to communication avoidance, and validate boundary decisions. However, community responses infrequently acknowledge distance contexts (13-18\%), or offer unhelpful abandonment advice. This created patterns of invisibility and misalignment between support sought and support received.

We now discuss four key challenges, articulating what differentiates remote from co-located caregiving contexts and why addressing remote-specific needs is essential. For each challenge, we identify implications for designing tools, platforms, or online communities, as well as for healthcare services and resources. %organization design 

\subsection{Remote monitoring and communication}
\subsubsection{Challenge}
Unlike co-located caregivers who can observe their loved ones in everyday life \cite{allan2020perspectives, yamashita2013understanding}, remote supporters must infer their loved ones' well-being through limited digital signals, such as voice patterns in phone calls, writing styles of text messages, or social media activity. Remote supporters developed a unique sensitivity to these digital cues in trying to detect crises, relapses, or safety concerns. 
However, when receivers are withdrawn or experience acute symptoms, supporters lose access to these signals, and without direct observation of the receivers, they cannot distinguish between normal unavailability and concerning silence. In such situations, supporters turn to online peers to seek experiential knowledge for interpreting receivers' behavior, in order to manage the uncertainty and risk. But since remote care in these communities is underrepresented, their situation is rarely acknowledged and they fail to receive adequate advice. 
 
As such, remote supporters remain to infer safety and well-being through inadequate channels. Without appropriate awareness mechanisms, they might miss critical warning signs or make unnecessary emergency interventions that damage trust and receivers' autonomy. Both outcomes harm the caregiving relationship and can be a source of heightened anxiety and hypervigilance for supporters \cite{gumley2022digital}. These findings suggest that effective remote caregiving requires diverse remote communication and monitoring approaches beyond the traditional text and voice channels.

\subsubsection{Implications}
One potential intervention is supporter-receiver relational systems that help supporters and receivers signal presence and care without the burden of messaging and voice calling. For example, remotely triggered colored lights that fade over time, allowing either party to say ``\textit{I'm thinking of you}'' with minimal effort \cite{gaver2023living}. Similarly, digital games that allow asynchronous turn-taking \cite{devasia2025partnership} enable receivers to signal presence without having to articulate emotional states or engage in a conversation. These approaches support intentional connection with indirect, low-pressure gestures to stay connected, and minimal demands on expressivity.

Alternatively, digital tools can include passive, automated signals that require no receiver effort, and offer the supporter with well-being awareness. For example, sharing general patterns of daily activity (``they've been active today'') through gentle visual or haptic cues without revealing specific details \cite{wenhart2025relatedness}. Similarly, wearable-based systems can share physiological data (e.g., heart rate, activity levels) as ambient, playful representations rather than raw numbers \cite{liu2021significant}.  These automated signals provide reassurance to supporters while preserving receivers' privacy and autonomy.

Finally, online peer communities can provide collective sense-making for understanding ambiguous signals that monitoring technologies cannot always determine. Platform designers could strengthen these knowledge-sharing spaces by organizing supporters' experiential knowledge into searchable repositories, consolidating personal strategies and informational resources by specific challenges (e.g., interpreting silence during withdrawal, crisis intervention from a distance) \cite{marshall2024understanding, wadden2021effect}.

\subsection {Coordinating care across distance}
\subsubsection{Challenge}
Our analysis revealed that during acute episodes, remote supporters contact crisis helplines for guidance on how to respond. Many reported not knowing how to evaluate the severity from afar and when to call for outside help. Consistent with prior work on distance caregiving \cite{horowitz2018long}, when information is limited or unavailable, remote supporters tend to infer and respond to the worst-case scenario. 

%Access to local crisis services information

Further, when calling for help, our findings demonstrate that remote supporters often depend on local intermediaries, such as police wellness checks or family members, to perform in-person interventions they cannot execute themselves. In such situations, they must coordinate with local formal services and informal supporters (family, friends, neighbors), the latter of which may lack mental health training or hold conflicting views about treatment. When coordination breaks down, supporters experience heightened distress and uncertainty about whether their loved one is receiving appropriate care, while receivers may face inappropriate interventions or delays in critical support.

This analysis revealed tensions among the wider caregiving network, beyond those that arise in the carer-receiver relationship \cite{rudnik_carejournal_2024}. Our findings show that remote supporters worry about local caregivers' burden and its ripple effects on the wider family \cite{cagle_long-distance_2012}, such as concerns about the impact of a sibling's condition on their in-person caregiving parent. A support system of multiple family caregivers must carefully collaborate \cite{nikkhah2024family}, and our findings suggest that coordination breakdowns and conflicts impact the caregiving network and as such, the quality of support provided. 
This highlights the need to understand and design for caregiving as a distributed network beyond the immediate, local caregivers.}

\subsubsection{Implications}

Systems, whether digital platforms, helplines, or healthcare services, could provide remote supporters with curated information about specific local crisis resources \cite{honary2018web, pendse2020like}, helping them identify appropriate interventions from a distance. Robust local crisis services for in-person interventions are critical for responding to various situations \cite{bei_barriers_2023}.  
Online communities could also serve as spaces for seeking urgent guidance. Platforms could facilitate immediate, synchronous interactions between supporters based on crisis contexts, such as experiences with involuntary hospitalization, or coordinating with specific types of crisis services.

Tools that allow remote supporters, local caregivers, and providers to collectively track behavioral patterns \cite{zehrung2024transitioning, murnane2018personal} could help make visible warning signs that remote supporters cannot observe directly. These tools could also provide structures for stakeholders to explicitly define and communicate their roles and responsibilities \cite{hsu2024dancing, herbst2025recommendations}.  Surfacing each caregiver's responsibilities and workload \cite{nikkhah2024family} could provide transparency that enables caregivers to negotiate more equitable task distribution, and could be extended to local and remote care based on what each one can contribute given their situation.

\subsection{Boundary management and self-preservation}
\subsubsection{Challenge}
For co-located caregivers, physical proximity creates constant caregiving demands and reduces opportunities for respite \cite{broady_how_2015}, making boundary-setting both practically difficult and emotionally fraught. Our analysis revealed that, given the distance, boundary management is experienced differently by remote supporters. While they have greater opportunity to prioritize their well-being compared to co-located caregivers, stepping back is accompanied by the perception of abandoning the receiver, and as such, guilt ensues. 

Anonymous online communities, such as Reddit, can offer validation for caregiving decisions without fear of judgment \cite{daynes2023online}, including decisions related to setting firm boundaries. This aligns with research on ``experiential validation'' \cite{andalibi2021sensemaking}, where connecting with those who share similar experiences helps achieve perceived normalcy and provides therapeutic benefits by reducing isolation \cite{speirs2023lived}. 
However, our findings suggest potential misalignment in how communities respond to these boundary-related posts. Community members sometimes advised discontinuing the caregiving relationship entirely, which may not align with remote supporters' needs for strategies to maintain the relationship and continue the caregiving efforts while protecting their own well-being.  Instead, the unique boundary management challenges that remote supporters experience did not seem to be understood by the wider community, mostly populated by local caregivers, who view caregiving as an "all or nothing" situation. 

\subsubsection{Implications}
Creating safe, non-judgmental spaces is essential for enabling vulnerable sharing on sensitive health topics \cite{lazar2019safe}, including for caregivers discussing boundary-setting experiences. Platforms could be designed to support experience sharing in ways that help supporters better articulate their unique boundary-setting experiences and guidance needs. Recent work suggests approaches like multimedia storytelling tools \cite{MashaVideo2018} or interactive prompts that help caregivers better articulate their needs for validation \cite{progga2023just, an2024easytell, jha2025bring}. One risk when sharing information even on anonymous platforms such as Reddit, is breaching privacy, especially that of the receivers \cite{gatos2021hci}. Design could address this through prompts that focus on caregiver emotions rather than receiver details, or by enabling sharing within small trusted peer groups. Platforms could also surface common boundary-setting strategies and validate stepping back as necessary for sustainable caregiving. Integrating caregivers' experiential knowledge about boundary management \cite{gui2017investigating} could help normalize self-preservation practices that remote supporters often feel guilty about. 

%Normalizing boundary-setting
Beyond technology design, mental health resources should explicitly acknowledge that remote supporters need permission and strategies for boundary-setting. This includes access to respite services and support structures that enable breaks from caregiving without guilt \cite{van2015stigma, tixier2016counting, sin2012understanding}. While remote supporters have more freedom to choose their level of engagement with the receiver, they have needs for validating their choices to recognize that self-preservation is a legitimate caregiving practice rather than giving up.

\subsection{Recognition and representation in online caregiver communities} 
\subsubsection{Challenge}
Our comparison of remote support posts with non-remote posts revealed distinct engagement patterns: they were fewer (6\% of the dataset), longer, with fewer upvotes, and received fewer and shorter comments (Table \ref{tab:remote_nonremote}). Moreover, comments on remote supporters' posts infrequently acknowledged the remote context (13-18\% across posting purposes), suggesting that remote supporters may not receive advice tailored to their distance-related challenges. This lack of recognition aligns with previous findings that remote caregivers receive inadequate advice in online forums \cite{lee2025support} and illustrates how platforms can inadvertently marginalize minority user groups \cite{augustaitis2021online}. This invisibility prevents remote supporters from finding others with similar distance-related experiences, limiting the peer support benefits that come from shared understanding and experiences \cite{pendse2019cross, saksono2023evaluating}.

\subsubsection{Implications}
In light of these findings, platforms could facilitate connections among supporters based on shared contextual factors such as distance-related experiences, caregiving relationships, or cultural backgrounds \cite{pendse2019cross, peng2021effects, fang2022matching}. This would allow remote supporters to find others who understand the distance-specific challenges, such as coordinating crises across states or countries, or managing guilt about being remote. Beyond individual matching, platforms can promote the visibility of remote caregiving contexts to other remote caregivers and to the community at large. For instance, allowing supporters to explicitly tag posts with remote contexts (e.g., ``supporting from another state'') and indicate the type of advise they seek could help commenters provide tailored advice rather than defaulting to co-located assumptions. Platforms could also feature remote caregiving experiences through dedicated spaces or highlighted stories, increasing visibility and recognition of distance caregiving as a legitimate form of support.

\section{Future Work}
Our study focused on investigating posting and response patterns through individual posts rather than examining long-term community participation. Future work could investigate how remote supporters engage with these communities over time, including factors influencing sustained engagement. Understanding long-term participation is particularly critical because our findings suggest remote supporters may not receive targeted support in these general communities, predominantly comprised of co-located caregivers. Further, it is important to engage directly with community members to understand their motivations, rather than relying only on the analysis of their posts.  

Additionally, our focus on Reddit, a public forum platform, represents only one type of online space where remote supporters seek help. Future research could explore how patterns differ between public forums like Reddit and more private, closed groups such as Facebook Groups or Discord servers. The different affordances of these platforms, including community size, privacy levels and moderation approaches, may play a role in how remote supporters share their experiences and exchange support. 

\section{Conclusion}
Mental illness affects not only individuals but also their family members and close friends, who often take responsibility for crisis situations involving suicidal behavior and acute episodes. When supporters provide care from afar, this presents unique challenges and needs that differ from co-located caregiving. Like other caregivers, they seek help from online communities, with needs related to their remote caregiving experiences, specifically around remote monitoring and care coordination. We also found that seeking validation represents an important motivation for posting, reflecting remote supporters' need for both reassurance of normalcy and practical insights from others who understand their unique circumstances. However, remote supporters received less community engagement and rarely received distance-specific advice, suggesting potential marginalization within these predominantly co-located caregiving experiences. Our findings highlight the need for online caregiver communities to better accommodate diverse experiences through resources and features that connect peers. As remote caregiving becomes increasingly common, understanding and supporting these experiences is essential for comprehensive mental health care.

%\section{Acknowledgments}

\begin{acks}
We would like to thank Angela Chiang, Kayla Ristianto, Johnny Wang  and Hoi Ki Jenny Wong for their contributions to this research.
We are also grateful to Dr. Drew Margolin, Dr. Elaine Wethington, and the reviewers for providing helpful feedback. 
\end{acks}

\bibliographystyle{ACM-Reference-Format}
\bibliography{supporter_discourse}

\appendix
%\newpage
\section{Codebook} \label{codebook}
\begin{table*}[t]
  \caption{Purpose / RQ1: Frequencies and paraphrased example quotes of identified purposes among 522 posts. Each post is assigned 1-2 purposes.}
  \label{tab:codebook_purpose}
  \begin{tabular}{>{\raggedright\arraybackslash}p{6cm}cc>{\raggedright\arraybackslash}p{7cm}}
    \toprule
    \textbf{Purpose} & \textbf{\#} & \textbf{\%} & \textbf{Paraphrased example quote} \\
    \midrule
    \textbf{Request guidance on supporting receiver:} The poster seeks advice on how to support someone experiencing mental health challenges, or make sense of the receiver's behaviors, especially when those behaviors are confusing, harmful, or emotionally distant. & 289 & 55\% & \textit{``Now that she wants to break up again, I can't tell if it's an episode because we're long-distance. I'm so worried and confused if it a episode or actually. Is there ever a way to know ?''}; \textit{``He admitted my texts sometimes stress him out, but when I offered to stop, he reassured me to continue. How do I be a present, supportive partner?''} \\
    \midrule
    \textbf{Express emotion and share caregiving experience:} The poster uses the post to express emotional exhaustion, share personal journey, offering new perspective & 175 & 34\% & \textit{`` I had to protect myself from all that toxic energy, but it breaks my heart that I've lost the sister I care so much about.''} ; \textit{``I'm trying but this relationship is draining and I don't think I can continue without snapping on her.}'' \\
    \midrule
    \textbf{Seek validation and shared experience:} The poster seeks relatable, shared experience to make sense of their situation. & 131 & 25\% & \textit{``Everything I find online just says the same things, so I'm hoping to get real advice from folks who've actually been through this.''} ; \textit{``Has anyone actually succeeded in helping their loved one achieve long-term independence and stability?''}  \\
    \midrule
    \textbf{Seek coping strategies for oneself:} The poster looks for strategies to manage own well-being while in caregiving role, e.g., regulating anxiety, avoiding burnout, or handling grief. & 91 & 17\% & \textit{``Can I really help my alcoholic sister who's spiraling, is there actually anything I can do to help her, and if not, how do I come to terms with that?''} ; \textit{``Do I wait and give him more time to get better mentally, or should I end things now? I don't want to waste my time or his.''}  \\
    \bottomrule
  \end{tabular}
\end{table*}

\begin{table*}[t]
  \caption{Response Topic / RQ2: Frequencies and paraphrased example quotes of identified response topics. The topics include Provide emotional support, Offer informational strategies, Promote supporter well-being  Share personal experiences, and boundaries, and Acknowledge remote context.}
  \label{tab:codebook_response}
  \begin{tabular}{>{\raggedright\arraybackslash}p{6cm}cc>{\raggedright\arraybackslash}p{7cm}}
    \toprule
    \textbf{Response Topic/Subtopic} & \textbf{\#} & \textbf{\%} & \textbf{Paraphrased example quote} \\
    \midrule
    \multicolumn{4}{c}{\textbf{Provide emotional support (75\%)}} \\
    \midrule
    \textbf{Provide emotional support}: Expressing acceptance, reassurance, validation, empathy or sympathy. & 359 & 69\% & \textit{``It really does suck. I'm also really sorry about your situation. It's frustrating and scary when it feels like you can't do anything to help them.''} \\
    \midrule
    \textbf{Provide companion support}: Offering solidarity, presence, or shared experience. Often including invitations for further contact. & 268 & 51\% & \textit{``Feel free to DM me if you want to vent/ chat, as I think we are in very similar situations unfortunately.''}; \textit{``You're not alone! This is not the first time this has happened to someone.''} \\
    \midrule
    \multicolumn{4}{c}{\textbf{Offer informational strategies (69\%)}} \\
    \midrule
    \textbf{Provide actionable advice for support-giving}: Gives practical, specific advice on how to support another person with a mental health issue. & 244 & 47\% & \textit{``Try not to take anything he says personally and to resist the urge to argue with him. Trying to argue with a delusion just reinforces it. Empathizing with the emotions the person feels about it tends to discharge the tension.''} \\
    \midrule
    \textbf{Share educational knowledge}: Offers factual, medical, or caregiving information not framed as personal experience. & 231 & 44\% & \textit{``He does not recognize you right now. Bipolar Disorder attacks the prefrontal cortex, and he is probably incapable of feeling anything about your most significant memories together.''} \\
    \midrule
    \textbf{Provide external resources}: Recommends learning resources (book, YouTube, websites), support group information etc. & 181 & 35\% & \textit{``Are you familiar with the LEAP method of communication (book \& Ted talk)? And the workbook for bipolar people?''}; \textit{``For resources, check out the NAMI website, particularly for any info they have on bipolar disorder and schizophrenia.''} \\
    \midrule
    \textbf{Suggest receiver seek professional help}: Encourages the person they're supporting to seek therapy or professional mental health services & 106 & 20\% & \textit{``You need to get him to hospital. Get a mental health team to do an assessment. Or get the police to do a wellness check.''} \\
    \midrule
    \multicolumn{4}{c}{\textbf{Promote supporter well-being and boundaries  (65\%)}} \\
    \midrule
    \textbf{Emphasize self-care}: Encourages the poster to prioritize their own well-being, such as therapy, hobbies, healthy routines or boundaries. & 283 & 54\% & \textit{``Your partner's emotions matter of course, but so do yours. Sometimes dealing with depressed people can make you forget that since you can get so caught up worried about their well-being.''} \\
    \midrule
    \textbf{Advise discontinuing support relationships}: Advises the poster to stop supporting to the person they are concerned, or to end the relationship altogether. & 158 & 30\% & \textit{``Cut your losses now. If he's unmedicated, it won't get better... Find someone stable''}; \textit{``Let go of them. Don't answer the calls unless you want to. You don't have to be your sisters caretaker.''} \\
    \midrule
    \multicolumn{4}{c}{\textbf{Share personal experience (64\%)}} \\
    \midrule
    \textbf{Share personal experience}: Shares a first-person account or personal anecdote of a similar situation to offer relatability or insight. & 333 & 64\% & \textit{``That's the hardest part of being a partner to a pwbp (person with bipolar) is that the switch can turn off anytime and they transform into a totally different person. It is important to recognise and understand that this is the bp working.''} \\
    \midrule 
    \multicolumn{4}{c}{\textbf{Acknowledge remote context (11\%)}} \\
    \midrule
    \textbf{Acknowledge remote context}: Comments on the challenges or limitations of supporting someone physically distanced. & 55 & 11\% & \textit{``It must be so hard, having to witness all this at a distance and not being able to physically intervene. Do you know what his local support network is like? Has he got any friends living nearer, any family members who could be supportive? Can you get in touch with any of them?''} \\
    \bottomrule
  \end{tabular}
\end{table*}

\begin{table*}[t]
  \caption{Themes / RQ3 (Part A): Frequencies and paraphrased example quotes of identified themes in posts. This part covers \textcolor{supportblue}{Support and coping mechanisms} and \textcolor{emotionalpurple}{Emotional toll}. {$\bigstar$} indicates themes unique to remote support.}
  \label{tab:codebook_theme_partA}
  \begin{tabular}{>{\raggedright\arraybackslash}p{6cm}cc>{\raggedright\arraybackslash}p{7cm}}
    \toprule
    \textbf{Theme} & \textbf{\#} & \textbf{\%} & \textbf{Paraphrased example quote} \\
    \midrule
    \multicolumn{4}{c}{\textcolor{supportblue}{\textbf{Support and coping mechanisms}}} \\
    \midrule
    \textcolor{supportblue}{\textbf{Monitor symptoms and treatment adherence:}} The poster observes or tracks receiver's mental health state (e.g., rest and nutrition; shifts in speech, energy, episode patterns, medication-taking behavior). & 167 & 32\% & \textit{``It was becoming clear that he had stopped taking his medication because his symptoms weren't getting better, they were actually getting worse.''}; \textit{``Her behaviour has changed significantly. She talks far more than usual, very quickly, and sometimes doesn't finish her lines of thought.''}\\
    \midrule
    \textcolor{supportblue}{\textbf{Prioritize own mental health:}} The poster prioritizes their own mental health over caregiving their partner/family, set boundaries, or recognizing the need to let go/end the relationship. & 163 & 31\% & \textit{``In the end, it's hard but you really have to know when to put yourself first and not feel guilty for it.''}; \textit{``I am simply tired and not willing to subject myself to negative treatment from anyone.''}\\
    \midrule
    \textcolor{supportblue}{$\star$ \textbf{Apply multiple digital channels and cues:}} The poster describe texting, calling, video chatting, observing social media activites for caregiving, or using features such as location-sharing to get hold of the person being supported & 110 & 21\% & \textit{``She takes her pills via video call with myself or one of my two siblings nightly, showing us each pill removed from its bottle and followed by food and water.''};\textit{``We live long distance and he doesn't drink around me. I can tell he's been drinking at least weekly because his speech is slurred when we talk over FaceTime.''} \\
    \midrule
    \textcolor{supportblue}{\textbf{Urge receiver to seek professional help:}} The poster actively encourages the person being supported to seek therapy, medication, or professional support. & 75 & 14\% & \textit{``I've been encouraging him to get professional help because his depression has been affecting our relationship.''}\\
    \midrule
    \textcolor{supportblue}{$\star$ \textbf{Coordinate with local/other supporters:}} The poster mentions working with others nearby the person being supported (e.g., family, friends) to support the person being supported. & 66 & 13\% & \textit{``Her boyfriend calls me every day with updates, and he says the hospital psychologist is working to get her on a consistent medication routine.''}  ; \textit{``She lives with my mom about an hour away, so I'm not around her much, but my mom has noticed things like her not sleeping well and obsessing over past trauma.''} \\
    \midrule
    \textcolor{supportblue}{$\star$ \textbf{Limit contact to protect well-being:}} The poster intentionally reduces or ends communication with their loved one, such as blocking contact or distancing themselves. & 58 & 11\% & \textit{``After putting up with decades of being constantly criticized—I'm talking 30 years of this—I finally reached my breaking point last year. Now we barely talk.''}\\
    \midrule
    \textcolor{supportblue}{$\star$ \textbf{Seek formal support services:}} The poster describes (considering) reaching out to emergency mental health teams, hotlines and police to address the receiver's needs. & 54 & 10\% & \textit{``I have talked him out 5 or more suicide attempts where recently I had to call the cops on him.''} \\
    \midrule
    \textcolor{supportblue}{\textbf{Plan for care-related discussion:}} The poster expresses intent to initiate a conversation with the person they support about their mental health or related behaviors. & 50 & 10\% & \textit{``I'd hoped to bring this up during this visit. I was planning to keep it gentle ('I love you and I'm worried about you') but want to be specific about the drinking rather than just vaguely mentioning health concerns.''}  \\
    \midrule
    \textcolor{supportblue}{$\star$ \textbf{Leverage physical distance:}} The poster may intentionally keep a distance from the person being supported. Physical distance helps the poster cope, detach, or maintain perspective. & 27 & 5\% & \textit{``I intentionally live far away to prioritize my own family. I do not want to be an arm's reach away to fix a problem (or get blamed for not showing up'').} \\
    \midrule
    \multicolumn{4}{c}{\textcolor{emotionalpurple}{\textbf{Emotional toll}}} \\
    \midrule
    \textcolor{emotionalpurple}{\textbf{Emotional toll:}} The poster explicitly expresses helplessness, anxiety, uncertainty, isolation, loneliness, and guilt that are associated with their caregiving role. & 378 & 72\% & \textit{``Being so far away has been really difficult for me.''}; \textit{``I had no idea this illness could trigger such a severe manic episode, so experiencing it firsthand has been really shocking. I feel completely lost about how to help her.''}\\
    \bottomrule
  \end{tabular}
\end{table*}

\begin{table*}[t]
  \caption{Themes / RQ3 (Part B): Frequencies and paraphrased example quotes of identified themes in posts. This part covers \textcolor{primaryred}{Primary stressor}. {$\bigstar$} indicates themes unique to remote support. (Continued)}
  \label{tab:codebook_theme_partB}
  \begin{tabular}{>{\raggedright\arraybackslash}p{6cm}cc>{\raggedright\arraybackslash}p{7cm}}
    \toprule
    \textbf{Theme} & \textbf{\#} & \textbf{\%} & \textbf{Paraphrased example quote} \\
    \midrule
    \textcolor{primaryred}{\textbf{Perceived symptom-driven/ unpredictable behavior:}} The person being supported behaving in inconsistent or confusing ways due to mental illness, such as sudden mood shifts, changes in personality. & 222 & 43\% &  \textit{``His mood changes are so unpredictable. Sometimes he's extremely depressed. Other times he's happy and acts completely normal for weeks. Then there are manic phases where he barely sleeps, tackles tons of projects, and has big plans for life changes.''}  \\
    \midrule
    \textcolor{primaryred}{$\star$ \textbf{Receiver avoid communication:}} The person being supported is unresponsive, disengaged, or stops replying to texts, calls, or conversations. & 164 & 31\% & \textit{``My long-distance boyfriend is going through depression and requested some time alone. I keep sending messages to let him know I'm still around and thinking of him, but he hasn't replied in over a month.''} \\
    \midrule
    \textcolor{primaryred}{$\star$ \textbf{Rejection or avoidance of caregiver support:}} The person being supported actively resists or avoids the poster's attempts to help, such as refusing to discuss their mental health, ignoring offers of assistance. & 151 & 29\%  & \textit{``They refuse to stop drinking and have declined all suggestions for help, including therapy, community support,  moving closer to us so we can at least keep an eye on them.''} \\
    \midrule
    \textcolor{primaryred}{\textbf{Non-adherence to treatment/advice:}} The person being supported ingores clinical advice or inconsistently follows through with treatment (e.g., stopped taking medicine, avoid going to therapy/doctors, relapse from drinking). & 150 & 29\% & \textit{``She drinks a box of wine on her bus ride home from work and pretends to be trying to stop drinking.''}; \textit{''It's even more challenging because she doesn't really believe her diagnosis, skips her appointments, and doesn't take her medication consistently.''}\\
    \midrule
     \textcolor{primaryred}{\textbf{Acute mental health episode or relapse:}} The poster explicitly describes a recent acute worsening or relapse of their loved one's mental health condition, such as manic/psychotic/depressive episodes, or relapse into alcohol use.  & 111 & 21\% &  \textit{``She's suffering from severe psychotic symptoms, but she refuses to acknowledge that she needs treatment.''}; \textit{``I just found out yesterday that a friend of mine has been diagnosed with bipolar disorder and has been going through a manic episode for the last three weeks.''}\\
    \midrule
    \textcolor{primaryred}{\textbf{Receiver being institutionalized:}} The person being supported is currently in a treatment facility, voluntarily or not. & 75 & 14\% & \textit{``He's seriously at risk of harming himself and is currently in his second psych hold since this episode started 3 months ago.''} \\
    \midrule
    \textcolor{primaryred}{\textbf{Refusal of professional help:}} The person being supported is untreated and declines therapy, and medication. & 71 & 14\% &  \textit{``I've recommended therapy and offered other ways to help, but he turns down everything I suggest.''} \\
    \midrule
    \textcolor{primaryred}{$\star$ \textbf{Information gaps:}} The poster describes how distance leaves them without the full picture of their loved one's condition, making it hard to monitor and respond effectively. & 36 & 7\%  & \textit{``My question is how can we tell, from afar and with no communication, when he is a danger to himself or others… I'm trying to gauge his state''}; \textit{`Since I live in another state, I have no way to monitor how much she's actually consuming at home.''}\\
    \midrule
   \textcolor{primaryred}{$\star$ \textbf{Separation initiated by receiver:}} The poster experiences involuntary separation when the care receiver moved out, or fled the area while demonstrating symptoms. & 36 & 7\% & \textit{``My ex broke up with me six months ago during a manic episode, during which she also moved to another state. She was only recently diagnosed and just started treatment.''}; \textit{``My husband with unmedicated bipolar disorder impulsively asked for separation and suddenly left, abandoning me by moving away.''}
    \\ 
    \bottomrule
  \end{tabular}
\end{table*}

\begin{table*}[t]
  \caption{Themes / RQ3 (Part C):  Frequencies and paraphrased example quotes of identified themes in posts. This part covers \textcolor{secondaryorange}{Secondary stressor}. {$\bigstar$} indicates themes unique to remote support. (Continued)}
  \label{tab:codebook_theme_partC}
  \begin{tabular}{>{\raggedright\arraybackslash}p{6cm}cc>{\raggedright\arraybackslash}p{7cm}}
    \toprule
    \textbf{Theme} & \textbf{\#} & \textbf{\%} & \textbf{Paraphrased example quote} \\
    \midrule
    \textcolor{secondaryorange}{\textbf{Negative communication from receiver:}} The person being supported uses hurtful, dismissive, or emotionally charged language, often out of character or distressing. & 119 & 23\%  &  \textit{``Dealing with their hostile behavior during this episode was really challenging. There were hateful text messages. Even before this episode, they had a pattern of treating me badly over the years, lying, manipulative, selfish, and negative behavior.''} \\
    \midrule
    \textcolor{secondaryorange}{\textbf{Illness impact on supporter–receiver relationship:}} The poster describes how the receiver's mental illness has changed the emotional closeness, communication patterns, or quality of the relationship, such as feeling distant, abandoned, or no longer recognized as the same person. & 95 & 18\% & 
    \textit{``My brother has changed dramatically over three years.  an empathetic, caring person into someone obsessed and hostile who attacks family members about things that NEVER happened.''} \\
    \midrule
    \textcolor{secondaryorange}{$\star$ \textbf{Concerns about impact on other family members:}} The poster expresses worry about how caregiving is affecting someone else, e.g., a parent or sibling  & 74 & 14\% & \textit{``What can I do to support my mom in supporting my sister so she feels more equipped and not just get completely consumed by this nightmare?''} \\
    \midrule
    \textcolor{secondaryorange}{\textbf{Feeling unappreciated as partner/families:}} The poster feels undervalued, unsupported, or unreciprocated in their role. & 47 & 9\% & \textit{``She doesn't ask about me, she doesn't think about me, she's just keen to tell me ALL about herself.''}\\
    \midrule
    \textcolor{secondaryorange}{$\star$  \textbf{Conflicts with other supporters:}} The poster describes tension, disagreement, or frustration with other family members or partners involved in caregiving. & 35 & 7\% & \textit{``His family tends to be confrontational and doesn't communicate well, which just makes everything worse. That's why I've been trying to help him as much as I can.''} \\  
    \bottomrule
  \end{tabular}
\end{table*}

\end{document}